\documentclass[aps, pra, 10pt, superscriptaddress, twocolumn, longbibliography, floatfix]{revtex4-1}
\usepackage[nointlimits]{amsmath}
\usepackage{graphicx,nicefrac}
\usepackage{subfigure}
\usepackage{color}
\usepackage{txfonts}
\usepackage{mathtools}
\usepackage{hyperref}
\usepackage{tabularx}
\hypersetup{colorlinks=true, citecolor=blue, linkcolor=blue}

\usepackage{physics}
\usepackage{amsfonts,amssymb}
\usepackage{array,bm,color}
\usepackage{epsfig,graphicx,nomencl,revsymb4-1,upgreek,url}
\usepackage{hyperref}
\usepackage{algpseudocode}
\usepackage{xspace}
\usepackage{tabularx}
\usepackage[normalem]{ulem}
\usepackage{changes}
\usepackage{enumitem}
\usepackage{booktabs}
\usepackage{mathtools}

\newcommand{\Id}{\mathbb{I}}

\def\bra#1{{\left\langle #1 \right|}}
\def\ket#1{{\left| #1 \right\rangle}}

\newcommand{\avg}{\textrm{avg}}

\newcommand{\comp}[2]{\tilde{\mathsf{#2}}} 
\newcommand{\exact}[2]{\mathsf{#2}} 
\newcommand{\noisyop}[1]{\Lambda}
\newcommand{\idealop}[1]{\mathcal{U}}
\newcommand{\approxop}[1]{\tilde{\mathcal{U}}}
\newcommand{\gate}[1]{\mathsf{#1}}

\begin{document}
\title{Scalable Full-Stack Benchmarks for Quantum Computers}

\author{Jordan Hines}
\thanks{jordanh@berkeley.edu}
\affiliation{Department of 
Physics, University of California, Berkeley, CA 94720}
\affiliation{Quantum Performance Laboratory, Sandia National Laboratories, Albuquerque, NM 87185 and Livermore, CA 94550}
\author{Timothy Proctor}
\thanks{tjproct@sandia.gov}
\affiliation{Quantum Performance Laboratory, Sandia National Laboratories, Albuquerque, NM 87185 and Livermore, CA 94550}

\begin{abstract} 
Quantum processors are now able to run quantum circuits that are infeasible to simulate classically, 
creating a need for benchmarks that assess a quantum processor's rate of errors when running these circuits.
Here, we introduce a general technique for creating efficient benchmarks from any set of quantum computations, specified by unitary circuits. Our benchmarks assess the integrated performance of a quantum processor's classical compilation algorithms and its low-level quantum operations. Unlike existing ``full-stack benchmarks'', our benchmarks do not require classical simulations of quantum circuits, and they use only efficient classical computations. 
We use our method to create randomized circuit benchmarks, including a computationally efficient version of the quantum volume benchmark, and an algorithm-based benchmark that uses Hamiltonian simulation circuits. We perform these benchmarks on IBM Q devices and in simulations, and we compare their results to the results of existing benchmarking methods.
\end{abstract}
\maketitle

\section{Introduction}

As quantum computers increase in size and capabilities there is a pressing need for benchmarking tools to assess their performance. Many benchmarking protocols for quantum computers, such as randomized benchmarking \cite{magesan2011scalable, magesan2012characterizing, proctor2021scalable, proctor2018direct, McKay2020-no, hines2022demonstrating, helsen2018new, Helsen2020-it, Claes2020-cy,  Helsen2020-mb, cross2016scalable, emerson2005scalable, emerson2007symmetrized} and cross entropy benchmarking \cite{liu2021benchmarking, boixo2018characterizing, arute2019quantum}, measure the performance of a quantum processor's gates. However, gate performance alone is insufficient to assess a processor's computational capabilities, as a processor's ability to run a particular algorithm also depends on characteristics such as device connectivity, and the performance of classical algorithms for routing and compilation. Motivated by this fact, benchmarks that measure the integrated performance of classical pre-processing and quantum operations have been developed---often referred to as ``full-stack benchmarks'' \cite{cross2018validating, Mills2021application}. 

\begin{figure*}
    \centering
    \includegraphics{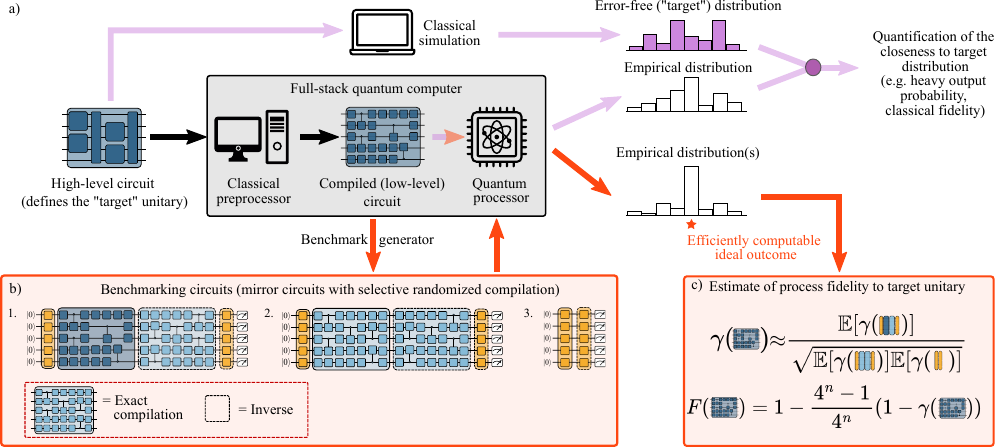}
    \caption{\textbf{A scalable full-stack benchmark generator.} (a) A full-stack quantum computer's performance on a target application is determined by the integrated performance of classical preprocessing and the quantum processor. Existing full-stack benchmarks (purple) consist of tasking a full-stack quantum computer with running high-level quantum circuits. The quantum computer first performs classical preprocessing, then runs the compiled (``low-level'') circuits on a quantum processor. To compute performance on the benchmark, the results are compared to a classical simulation of the high-level circuits, which is typically exponentially expensive to perform. Our full-stack benchmarking method uses a different approach: it uses the compiled circuits to generate efficiently verifiable benchmarking circuits, which are then run on the quantum processor. (b) These benchmarking circuits are  mirror circuits of three types that use a compiled circuit output by the full-stack quantum computer's classical preprocessing and a pre-compiled exact compilation of the target circuit. (c) The results of running these circuits are used to efficiently compute the process fidelity of the processor's imperfect implementation of the target unitary.}
    \label{fig:protocol}
\end{figure*}

Full-stack benchmarks task an integrated (a.k.a.~``full-stack'') quantum computer with running a set of $n$-qubit quantum circuits specified using a hardware-agnostic gate set (a.k.a.~``high-level'' circuits). 
Full-stack benchmarks are increasingly being used to measure progress in quantum processor performance \cite{chen2023benchmarking, moses2023race}. For example, the quantum volume benchmark---a randomized, average-performance full-stack benchmark---has been widely adopted \cite{cross2018validating, Pelofske2022, Baldwin2022reexaminingquantum, miller2022improved, Moll_2018, amico2023defining}, and there are an increasing number of application-inspired benchmarks focused on measuring task-specific performance \cite{Mills2021application, 
 lubinski2023applicationoriented}. 
However, there is a critical limitation in the methodology behind existing full stack-benchmarks (Fig.~\ref{fig:protocol}, purple).

In existing full-stack benchmarks (i) the processor uses classical compilation and routing algorithms to turn the benchmark's ``high-level'' circuits into hardware-specific circuits, (ii) the processor runs those circuits on the processor's qubits, and then (iii) the quantum processor's output is compared to a classical simulation, to quantify the error in the implementation (see Fig.~\ref{fig:protocol}). Some existing full-stack benchmarks use high-level $n$-qubit circuits that are efficiently classically simulatable, and others do not, but both of these scenarios cause problems for the benchmark. If a benchmark's high-level circuits cannot be efficiently simulated classically, the benchmark requires exponentially expensive classical computations [in step (iii), above] and is therefore not scalable. This is the case with, for example, the quantum volume benchmark. If a benchmark's high-level circuits can be efficiently simulated, the benchmark's analysis is scalable, but the results may not accurately capture the full-stack processor's performance on similar circuits that cannot be simulated classically. For example, one approach to constructing a scalable benchmark from a class of parameterized circuits (e.g. Hamiltonian simulation circuits) is to choose a subset of there circuits that are easy to simulate (e.g., Clifford circuits). However, a processor's compilation algorithms may find more efficient compilations for these efficiently simulable circuits than for others in the same circuit class. Reliable, scalable full-stack benchmarking therefore requires a new approach.

In this paper, we introduce a method for creating robust and efficient full-stack benchmarks from any set of unitary circuits. In our approach (Fig.~\ref{fig:protocol}, orange), high-level circuits are sent to the full-stack processor for classical pre-processing, but then the resulting compiled circuits are used as \emph{subroutines} in the benchmarking circuits ultimately run on the quantum processor. These benchmarking circuits have an efficiently computable error-free outcome, and their results can be used to efficiently estimate the process fidelity with which the processor performed the target unitary operation \cite{proctor2022establishing}.  
Our method can be used to efficiently and reliably test progress towards implementing quantum algorithms, and it enables precise assessment of processor performance on classically infeasible computation. We detail the use of our method to create scalable benchmarks with three types of circuits: quantum volume circuits, random circuits defined with grid and linear geometries, and Hamiltonian simulation circuits. We demonstrate these three benchmarks in simulations and on IBM Q devices. 

\section{Full-Stack Benchmarking Preliminaries}

We start by reviewing the concept of full-stack benchmarking. A full-stack quantum computer consists of (i) a classical processing stack that can, e.g., compile quantum circuits, and (ii) a quantum processor (qubits and quantum logic gates) that can execute ``low-level'' quantum circuits containing only that processor's native quantum gates. Many existing quantum benchmarking protocols (e.g., RB \cite{magesan2011scalable, magesan2012characterizing} and cross-entropy benchmarking \cite{liu2021benchmarking, boixo2018characterizing, arute2019quantum}) bypass a processor's classical compilation routines and run low-level circuits directly on the quantum processor, allowing them to measure the performance of native gates directly.  
In contrast, full-stack benchmarks measure the integrated performance of a full-stack quantum computer by tasking the full-stack quantum computer with running high-level circuits (see Fig.~\ref{fig:protocol}), which must undergo classical preprocessing to convert them to low-level circuits that can be run on the quantum processor.

A full-stack benchmark tasks a full-stack processor with implementing one or more unitaries $U$. Each unitary is specified by a ``high-level'' quantum circuit $\gate{C}_{\textrm{high-level}}$, i.e., a circuit that has not been compiled for a particular device connectivity or native gate set. The processor's classical pre-processor must map each high-level circuit $\gate{C}_{\textrm{high-level}}$ onto its quantum processor, producing a low-level circuit $\comp{U}{C}$ that, when run on the quantum processor, implements a quantum process $\Lambda$ (which is a completely positive and trace preserving map). A full-stack processor's task when running a full-stack benchmark is to implement a process $\Lambda$ that is as close as possible to $\mathcal{U}$ (or, as we discuss below, any of a set of closely related unitaries), i.e., the goal is to minimize the difference between $\Lambda$ and $\mathcal{U}$, where $\mathcal{U}[\rho]=U\rho U^{\dagger}$ is the superoperator representation of $U$.
Note that the circuit $\comp{U}{C}$ is permitted to be an \emph{approximate} compilation of $\gate{C}_{\textrm{high-level}}$, i.e., in the absence of errors it implements a unitary $\tilde{U}$ such that $\tilde{U} \neq U$ (although to perform well on the benchmark it is necessary for $\tilde{U} \approx U$). To achieve the best possible implementation of $U$, the full-stack processor must balance approximation error in circuit compilation with the quantum processor's gate errors.

A full-stack benchmark's task is often slightly different from implementing a unitary $U$. Instead the task is to implement $U$ up to operations that can be accounted for in classical post-processing of the data. Formally, the task is to implement any $U'$ satisfying $U'=VU$, where $V \in \mathbb{V}$, and $\mathbb{V}$ is a set of unitaries specified by the benchmark. For example, $\mathbb{V}$ could be the set of all unitaries corresponding to permutations of the qubits, i.e., the task is to approximation $\mathcal{P}\mathcal{U}$, where $\mathcal{P}$ is the superoperator representation of any qubit permutation operator. The specific benchmarks we implement in this work use this task, but our method can also create benchmarks with the more stringent task of approximating $\mathcal{U}$ (or a less stringent task corresponding to a larger $\mathbb{V}$). We allow qubit permutations because it captures a common compiler optimization: a full-stack processor may use simple classical post-processing to permute final measurement result. In devices with limited qubit connectivity, this optimization often reduces the need for high-error SWAP networks.

Existing full-stack benchmarks  (Fig.~\ref{fig:protocol}, purple) run each low-level circuit $\comp{U}{C}$ on the quantum processor and compare the observed output distribution to the error-free output distribution of $\gate{C}_{\textrm{high-level}}$. However, computing the error-free output distribution of $\gate{C}_{\textrm{high-level}}$ typically requires classical computation cost that scales exponentially in the number of qubits. 
The benchmarks we introduce in this work use a different metric of performance, the \emph{process fidelity},
\begin{equation}
    F(\noisyop{U}, \mathcal{U}) = \frac{1}{4^n}\Tr(\mathcal{U}^{\dag}\noisyop{\comp{U}{C}}).
\end{equation}
We will often find it more convenient to use a rescaling of the fidelity called the \emph{polarization}:
\begin{equation}
        \gamma(\mathcal{U}^{\dag}\noisyop{\comp{U}{C}}) = \frac{4^n}{4^n-1} F(\mathcal{U}^{\dag}\noisyop{\comp{U}{C}}) -  \frac{1}{4^n-1}.  \label{eq:pol_def}
\end{equation}

\section{Mirror Full-Stack Benchmarking Method}
\label{sec:method}

In this section we introduce our method for creating efficient and scalable full-stack benchmarks, which is summarized in Fig.~\ref{fig:protocol} and described in more detail in Appendix~\ref{app:method}. Our method is a \emph{benchmark generator}---i.e., given a set of high-level circuits, it constructs a benchmark based on those circuits. The resulting benchmark measures the process fidelity with which a full-stack quantum computer performs a unitary $U'=PU$, where $U$ is the target unitary encoded by high-level circuit $\gate{C}_{\textrm{high-level}}$ and $P$ is a unitary corresponding to a known permutation of the qubits (this permutation is specified as part of  the output of the classical preprocessor). The key insight of our method is that we can avoid simulating $\gate{C}_{\textrm{high-level}}$ (or its compilation $\comp{U}{C}$) to verify the output of our benchmark by using $\comp{U}{C}$ as a component of the benchmarking circuits, rather than running $\comp{U}{C}$ in isolation on the quantum processor. 
Our protocol is the following:
\begin{enumerate}
\item[(i)] Use the full-stack processor's classical pre-processing to compile each $\gate{C}_{\textrm{high-level}}$.
\item[(ii)] For each compiled circuit $\comp{U}{C}$, run an ensemble of circuits (detailed below) that are built from $\comp{U}{C}$ and that enable efficient estimation of the fidelity of the noisy implementation of the compiled circuit [$\Lambda$] to $\mathcal{U}'$. 
\item[(iii)] Estimate the fidelity $F(\Lambda,\mathcal{U}')$ for each $\gate{C}_{\textrm{high-level}}$ from the data.
\end{enumerate}

Our protocol estimates the fidelity of $\Lambda$ to $\mathcal{U}'$ [steps (ii) and (iii) above] by adapting a recently-introduced scalable fidelity estimation protocol: mirror circuit fidelity estimation (MCFE) \cite{proctor2022establishing}. MCFE is a method that, given (1) a ``test circuit'' $\gate{C}$ and (2) a ``reference circuit'' $\gate{C}_{\textrm{ref}}$ that contains only single-qubit gates and self-inverse Clifford two-qubit gates (e.g., CNOT or CPHASE), efficiently estimates the fidelity of a quantum processor's implementation of $\gate{C}$ to the unitary evolution specified by $\gate{C}_{\textrm{ref}}$. MCFE achieves this by running an ensemble of mirror circuits \cite{proctor2020measuring} constructed from $\gate{C}$ and $\gate{C}_{\textrm{ref}}$. These circuits enable estimating the process fidelity in a way that is robust to error in the processor's implementation of $\gate{C}_{\textrm{ref}}$ and state preparation and measurement error. 

We could use MCFE to estimate the fidelity of $\Lambda$ to $\mathcal{U}'$ by using the compiled circuit $\comp{U}{C}$ as the test circuit and an \emph{exact} compilation of $U'$ as the reference circuit. Instead, we make a slight modification to MCFE that decreases the overhead incurred by running $\gate{C}_{\textrm{ref}}$ as a subroutine: our method only requires that $\gate{C}_{\textrm{ref}}$ is an exact compilation of $U'$ up to right multiplication by a permutation of the qubits (see Appendix~\ref{app:method} for details). 

Our method runs mirror circuits of three types, shown in Fig.~\ref{fig:protocol}(b).
Each mirror circuit has a \emph{target bit string}, which is determined efficiently during circuit construction without full simulation of the circuit. If the full-stack processor implements $\gate{C}_{\textrm{high-level}}$ perfectly---i.e., $\Lambda = \mathcal{U}'$---then each mirror circuit always outputs the target bit string. For each mirror circuit $\gate{m}$, we estimate its \emph{observed polarization}, 
\begin{equation}
\gamma(\gate{m}) =  \frac{4^n}{4^n-1}\left[\sum_{k=0}^{n} \left(-\frac{1}{2}\right)^k h_k\right] - \frac{1}{4^n -1},
\label{eq:S}
\end{equation}
where $h_k$ is the probability that the circuit outputs a bit string with Hamming distance $k$ from its target bit string. Using the average observed polarization of $M \gg 1$ mirror circuits of each of the three circuit types, we estimate the fidelity of $\Lambda$ to $\mathcal{U}'$ using the equation 
\begin{equation}
    \hat{F}(\Lambda, \mathcal{U}') = 1 - \frac{4^n-1}{4^n}\left(1 - \frac{\avg_{\gate{m}_1}(\gamma(\gate{m}_1))}{\sqrt{\avg_{\gate{m}_2}(\gamma(\gate{m}_2))\avg_{\gate{m}_3}(\gamma(\gate{m}_3))}}\right), \label{eq:fidelity_estimate}
\end{equation}
where $\gate{m}_j$ denotes a mirror circuit of type $j$ [see Fig.~\ref{fig:protocol}(b-c)]. The theory of MCFE (see \cite{proctor2020measuring}) implies that $\hat{F}(\Lambda, \mathcal{U}')$ is a reliable estimate of $F(\Lambda, \mathcal{U}')$, even when no operations are assumed to be perfect. Furthermore, our method is sample efficient, as the number of samples required to obtain a fixed estimation precision does not grow with $n$. However, we note that the number of circuits required to obtain a reliable estimate of $\hat{F}(\Lambda, \mathcal{U}')$ grows as $F(\Lambda, \mathcal{U}')$  decreases \cite{proctor2020measuring}. 

\section{Scalable Full-Stack Random Circuit Benchmarks}\label{sec:random_circuits}
In this section, we use our benchmark generator to define scalable randomized full-stack benchmarks. We first introduce a scalable quantum volume benchmark, then introduce benchmarks that use circuits defined on grid and linear geometries. 

\subsection{Mirror Quantum Volume}
\label{sec:mqv}
The quantum volume (QV) benchmark \cite{cross2018validating} is a widely-used randomized full-stack benchmark that produces a single-number metric for the capabilities of a quantum processor. The QV protocol requires simulating $n$-qubit random circuits to determine their \emph{heavy outputs}. This classical computation's time cost scales exponentially in $n$. In this section, we apply our benchmark generator to QV circuits to eliminate the need for expensive classical simulation.

\subsubsection{Quantum Volume Preliminaries}

We start by reviewing the standard quantum volume benchmark. An $n$-qubit quantum circuit $\gate{C}_{\textrm{QV}}$ with benchmark depth $d$ consists of $d$ layers of $\lfloor \nicefrac{n}{2} \rfloor$ random SU(4) gates on a random pairing of the $n$ qubits (for odd $n$, a single qubit idles in each layer). Success on a QV circuit is measured by the full-stack processor's ability to produce the $2^{n-1}$ most likely measurement outcomes of $\gate{C}_{\textrm{QV}}$, which are called \emph{heavy outputs}, at close to the same rate as an error-free implementation of $\gate{C}_{\textrm{QV}}$. Let $p_{\gate{C}_{\textrm{QV}}}(x)$ be the probability of measuring $x \in \{0,1\}^n$ when $\gate{C}_{\textrm{QV}}$ is run in the absence of errors. The set $H$ of heavy outputs of a quantum circuit is the set of bit strings that are measured with above median probability, i.e.,
\begin{equation}
    H = \{ x \mid p_{\gate{C}_{\textrm{QV}}}(x) > p_{\textrm{med}}\},
\end{equation}
where $p_{\textrm{med}}$ is the median of $\{p_{\gate{C}_{\textrm{QV}}}(x)\}_{x \in \{0,1\}^{n}}$. The QV benchmark consists of running $K$  quantum volume circuits of a fixed shape and computing their average \emph{heavy output probability} $\tilde{p}_H$, the probability that the processor outputs a bit string $x \in H$, averaged over all $K$ circuits (typically $K > 100$).  
Computing the heavy output probability requires determining the heavy outputs of each circuit, which requires full simulation of the error-free QV circuits. Due to this expensive classical computation, QV is computationally infeasible on a large number of qubits ($n \gtrsim 50$).

The standard QV benchmark is often used to determine a full-stack processor's \emph{quantum volume}. The quantum volume is $2^q$, where $q$ is the largest number such that width $q$, depth $q$ QV circuits achieve a heavy output probability of $\tilde{p}_H > \nicefrac{2}{3}$ with at least $97.5 \%$ confidence.

\subsubsection{Mirror quantum volume protocol}

Here, we present our \emph{mirror quantum volume} benchmark (MQV), a scalable version of quantum volume that consists of applying our benchmark generator to quantum volume circuits. A MQV experiment with width $n$ and benchmark depth $d$ is performed as follows:
\begin{enumerate}
    \item  Generate $K$ $n$-qubit QV circuits of depth $d$, i.e., circuits with $d$ benchmarking layers, each chosen by (1) generating a random pairing of the $n$ qubits (with a random qubit left idle when $n$ is odd), and (2) applying Haar-random SU(4) gates to each pair.
    \item Send each circuit to the full-stack processor's pre-processing to get a compiled circuit $\comp{U}{C}$. Additionally, compute an exact compilation $\exact{U}{C}_{\textrm{ref}}$ (up to qubit permutation) that contains only single-qubit gates and self-inverse Clifford two-qubit gates. 
    \item Use $\comp{U}{C}$ and $\exact{U}{C}$  to construct a set of benchmarking circuits (using the algorithm of Fig.~\ref{fig:protocol}b). Run these circuits, then estimate the average polarization of each QV circuit [Eq~\eqref{eq:fidelity_estimate}, Fig.~\ref{fig:protocol}c]. 
    \item Compute the average polarization $\bar{\gamma}_{n,d}$ of all $K$ QV circuits. 
\end{enumerate}
The result of an MQV experiment is an estimate $\bar{\gamma}_{n,d}$ of the average polarization of width $n$, depth $d$ QV circuits performed on the full-stack processor. Additionally, the results of our protocol can be used to infer the quantum volume of the processor: The quantum volume of the processor is approximately $2^n$, where $n$ is the largest value such that 
\begin{equation}
    \bar{\gamma}_{n,n}> \frac{1}{3\ln(2)} \label{eqn:qv_pass_pol}
\end{equation}
with at least $97.5\%$ confidence.

A version of QV using mirror circuits \emph{without} randomized compilation has been proposed as a heuristic for estimating standard QV results \cite{amico2023defining}. That method is computationally efficient, as every circuit ideally implements the identity operation. However, its reliability for quantifying performance on QV circuits is not guaranteed, because a quantum processor's performance on those circuits is highly dependent on the type of error present in the device (these circuits are a form of Loschmidt echo, and therefore susceptible to systematic cancellation or addition of coherent errors). In contrast, our MQV protocol's reliability is supported by the theory of MCFE \cite{proctor2022establishing}. Note, however, that the protocol of Ref.~\cite{amico2023defining} requires significantly fewer circuits than our method.
\begin{figure}
    \centering    
    \includegraphics{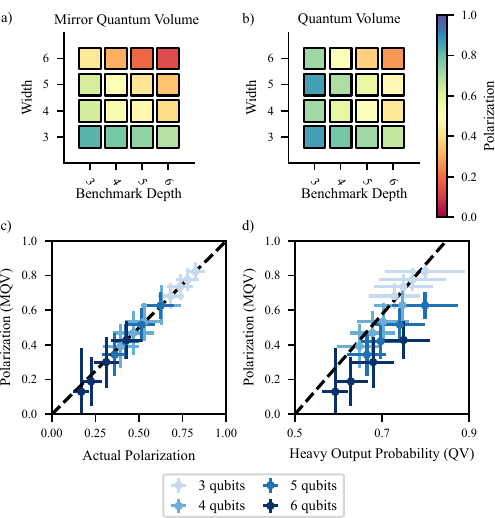}
    \caption{\textbf{Simulations of MQV.} We simulated MQV and the standard quantum volume protocol on $n=3,4,5,6$ qubits with single- and two-qubit depolarizing error. (a) The mean polarization of random SU(4) circuits, measured by MQV. (b) The observed heavy output probability, rescaled to an estimated polarization, of the QV circuits. (c) We compare the polarization estimates from MQV to the exact polarizations of the compiled circuits. (d) The heavy output probability from the QV benchmark versus the estimated polarization from our method. The average heavy output probability typically overestimates the polarization of the compiled circuits.}
    \label{fig:mqv_ibmq}
\end{figure}
\subsubsection{Theory}
MQV measures the average polarization of QV circuits, whereas the standard quantum volume benchmark measures their average heavy output probability. We now derive an approximate relationship between the polarization and the heavy output probability of QV circuits, justifying the condition given by Eq.~\eqref{eqn:qv_pass_pol}. 
We express the processor's error in implementing $\gate{C}_{\textrm{QV}}$ as a post-circuit error channel $\mathcal{E}$---i.e., $\Lambda = \mathcal{E}\mathcal{U}(C)$.
We approximate $\mathcal{E}$ by a global depolarizing error channel, 
\begin{equation}
    \mathcal{E}[\rho] = \gamma\rho + \frac{(1-\gamma)}{2^n} \mathbb{I}.
\end{equation}
We now relate the polarization $\gamma$ to the probability of measuring a heavy output upon performing $\Lambda$. Let $\rho_{\textrm{ideal}}=U\ketbra{0}{0}U^{\dag}$ be the state produced by performing the target unitary perfectly.
The probability of measuring a heavy output upon performing $\Lambda$ is
\begin{align}
    \tilde{p}_H & = \sum_{x \in H}\Tr(\ket{x}\bra{x} \mathcal{E}[\rho_{\textrm{ideal}}]) \\
    & = \sum_{x \in H}\left(\gamma \Tr(\ketbra{x}\rho_{\textrm{ideal}}) + \frac{1-\gamma}{2^n}\Tr(\ketbra{x}\mathbb{I})\right) \\
    & = \gamma p_H + \frac{1-\gamma}{2}, \label{eq:ideal_ph}
\end{align}
where $p_H$ is the probability of measuring a heavy output when performing the target unitary. We can approximate $p_H$ based on the distribution of the outcomes of random SU(4) circuits in the large-$d$, large-$n$ limit, which is used  to motivate the QV pass cutoff $\tilde{p}_H > \nicefrac{2}{3}$ \cite{cross2018validating}. In the large-$d$ limit, the error-free output probability distribution approaches the Porter-Thomas distribution \cite{aaronson2016complexitytheoretic, boixo2018characterizing}, for which the probability of a heavy output is $p_H = \nicefrac{(1 + \ln 2)}{2}  \approx 0.85$ \cite{aaronson2016complexitytheoretic}. Using this approximation for $p_H$, Eq.~\eqref{eq:ideal_ph} becomes
\begin{align}
    \tilde{p}_H = \frac{1 + \gamma \ln 2}{2}. \label{eq:tilde_ph}
\end{align}

The condition for passing the $n$-qubit QV test is that the average observed heavy output probability of $n$-qubit, depth $n$ QV circuits is $\tilde{p}_H> \frac{2}{3}$. Substituting this into Eq.~\eqref{eq:tilde_ph}, we obtain $\gamma > \frac{1}{3 \ln 2}$, which is the condition given in Eq.~\eqref{eqn:qv_pass_pol}.
More precisely, the QV success condition is that the average heavy output probability is greater than $\nicefrac{2}{3}$ with sufficiently high confidence, typically $97.5\%$ ($2\sigma$). Using the relation between $\tilde{p}_H$ and $\gamma$ given in Eq.~\eqref{eq:tilde_ph}, it is straightforward to write the success condition initially proposed for QV in \cite{cross2018validating}, which uses a normal approximation to obtain a simple analytic form for the confidence interval, in terms of $\gamma$. Alternatively, numerical confidence intervals on the polarization can be computed.

The theory presented above enables using the results of MQV to estimate the quantum volume of a full-stack processor. However, the relationship between heavy output probability and polarization in Eq.~\eqref{eq:tilde_ph} is approximate, and not accurate for all circuit shapes. This is because it is based on assuming Porter-Thomas statistics for the output probability distribution of QV circuits, which is not valid for small $d$ or small $n$. Even standard square (i.e., depth $n$) QV circuits are not deep enough to reach the regime where their outputs are accurately reflected by Porter-Thomas statistics \cite{Baldwin2022reexaminingquantum}. See Appendix~\ref{app:mqv} for further discussion. We note, however, that determining the quantum volume of a processor is not the main goal of our MQV benchmark---MQV can provide more detailed information about a processor than just its quantum volume, and its process fidelity estimates are reliable for all circuit shapes.

\subsubsection{Simulations}
We simulated MQV on $n = 3, 4, 5, 6$ qubits with circuits of benchmark depth $d = 3,4,5,6$. We simulated a quantum processor with heavy-hexagon geometry with $X_{\pi/2}$, $Z(\theta)$, and $\textrm{CNOT}$ native gates. Each single-qubit gate had single-qubit depolarizing noise with a polarization of $\gamma_1=0.9999$, and each two-qubit gate had two-qubit depolarizing noise with a polarization of $\gamma_2=0.975$. These simulations used a classical preprocessing routine in \texttt{Qiskit} \cite{Qiskit} that does no approximation in compilation, so that each compiled circuit $\comp{U}{C}$ is an exact compilation of $\gate{C}_{\textrm{QV}}$. We generated 2000 QV circuits of each shape $(n,d)$ and estimated the polarization of each QV circuit using MQV with $50$ of each of the three types of mirror circuits used in our method (see Fig.~\ref{fig:protocol}b). We also simulated standard QV with the same 2000 QV circuits---i.e., we simulated the compiled QV circuits and computed their observed heavy output probability. 

Figure~\ref{fig:mqv_ibmq} shows the results of these simulations. Figure~\ref{fig:mqv_ibmq}a shows the average polarization of circuits of each shape, as estimated by MQV. In Figure~\ref{fig:mqv_ibmq}c we compare the polarization estimates from MQV to the exact circuit polarizations.  The MQV polarization estimates agree with the exact circuit polarizations, demonstrating that MQV is accurate.

We also compare the results of MQV to the results of standard QV. To make this comparison, we convert the average heavy output probability of each QV circuit to an estimated polarization using Eq.~\eqref{eq:tilde_ph}. We plot these rescaled heavy output probabilities in Fig.~\ref{fig:mqv_ibmq}b, and we compare the rescaled heavy output probability to the MQV-estimated polarizations in Fig.~\ref{fig:mqv_ibmq}d. Typically, the effective polarization estimated from the heavy output probability of QV circuits is higher than the MQV estimate of the circuit polarization, with a greater discrepancy for larger $n$. For some circuit shapes, particularly shallow, low-width circuits, the error-free heavy output probability is larger than $0.85$, and as a result the heavy output probability sometimes overestimates the polarization in this regime.

\begin{figure*}
    \centering
    \includegraphics[scale=0.8]{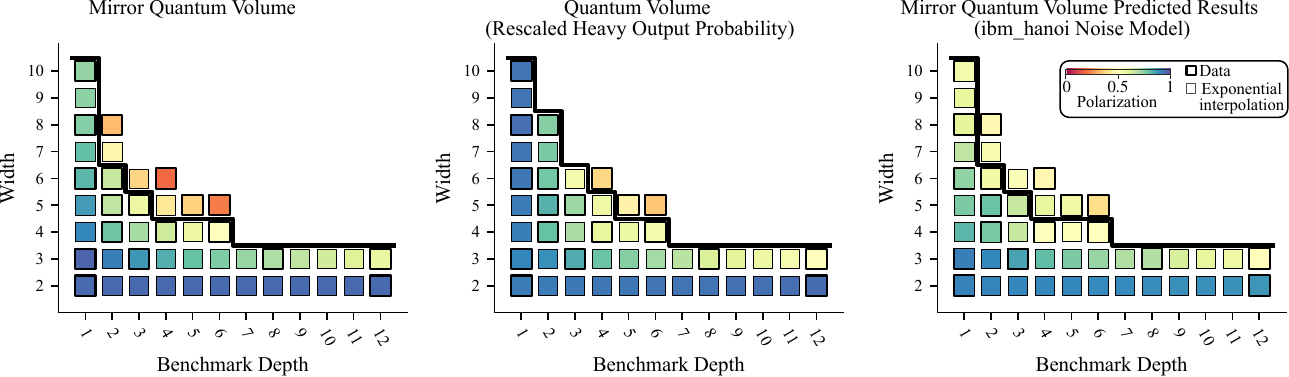}
    \caption{\textbf{Demonstrating MQV on IBM Q.} We ran the MQV and QV benchmarks on \texttt{ibm\_hanoi}. We plot the estimated average polarization of the QV circuits from our benchmark (left plot), and we compare these results to those of the QV benchmark with the same high-level circuits (center plot) and to predictions of our benchmark's results using a noise model based on the processor's calibration data (right plot). Because we did not run our benchmark using an exhaustive set of circuit shapes, we use an exponential interpolation of the data (dark-outlined boxes) to estimate the average polarization of QV circuits of other shapes (light-outlined boxes). We fit the average polarization data for a fixed circuit depth to  $\bar{\gamma}_{n,d} = Ap^d$ and use these exponential fits to estimate $\bar{\gamma}_{n,d}$ for depths not included in our experiment. For circuit shapes with $n > 6$ or $d=1$, we instead interpolate by fitting an exponential $\bar{\gamma}_{n,d} = Ap^n$ to all data of a fixed benchmark depth $d$. The black line indicates where the average polarization falls below the QV pass threshold, $\bar{\gamma}_{n,d} \approx 0.48$.}
    \label{fig:mqv_exp_ibmq}
\end{figure*}

\subsubsection{Demonstration on IBM Q}
\label{sec:mqv_demo}
We ran MQV and QV on \texttt{ibm\_hanoi}, an IBM Q processor with heavy-hexagon connectivity, with circuits of width up to $n=10$ and benchmark depth up to $d=12$. For each of $18$ circuit shapes, we generate $20$ QV circuits. We use these circuits as the high-level circuits for our benchmark, and generate $25$ of each of the three types of mirror circuits required by our method. In these demonstrations, we specify the classical preprocessing using $\texttt{Qiskit}$. The classical preprocessing we use produces exact compilations of the high-level circuits (up to qubit permutation), and it adds X-X dynamical decoupling on idling qubits during two-qubit gates to reduce the impact of coherent errors \cite{viola1999dynamical}. 

We also ran the compiled QV circuits, and we perform a heavy output probability analysis on these circuits to compare QV to MQV. For each QV circuit, we compute its error-free heavy output probability $p_H$ and the observed heavy output probability $\tilde{p}_H$. We then compute an estimated circuit polarization using Eq.~\eqref{eq:tilde_ph}. This is slightly different from standard QV analysis, which only analyzes the average heavy output probability over all QV circuits of a fixed shape. We choose this analysis because for low-depth circuits, the error-free heavy output probability may vary significantly between circuits, and it is frequently not close to the standard QV ``success'' threshold $p_H = \nicefrac{(1+\ln 2)}{2}$ (See Appendix~\ref{app:mqv} for the results of analysis based on the standard QV threshold).

Figure~\ref{fig:mqv_exp_ibmq} shows the results of our MQV and QV experiments in volumetric benchmarking plots \cite{blume2019volumetric}. We estimate the average polarizations for additional circuit shapes using an exponential decay heuristic (light-outlined boxes). The rescaled heavy output probability is systematically greater than the MQV-estimated polarizations, which we conjecture is due to heavy output probability being insensitive to $Z$-type errors immediately before measurement. 
 
We also compare the results of MQV to a prediction of the results of MQV from simulations using an error model based on the calibration data from the device at the time data was taken. The error model accurately predicts the MQV results for low-depth circuits and low-width circuits. We observe discrepancy at higher depths and widths, where errors not captured by the error model, such as coherent error and crosstalk error, are more likely to impact performance. 

\subsection{Demonstrating general randomized full-stack benchmarks}
\label{sec:randomized_demos}

\begin{figure}
    \centering
    \includegraphics{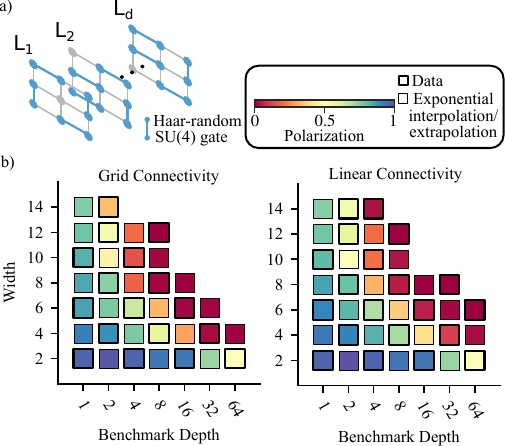}
    \caption{\textbf{Randomized full-stack benchmarks on IBM Q.} We ran random circuit full-stack benchmarks using high-level circuits defined on grid and linear geometries on \texttt{ibm\_hanoi}. (a) To generate a benchmark depth-$d$ high-level circuit for our grid geometry benchmark, we generate random circuits where each layer consists of $\lfloor \nicefrac{n}{2} \rfloor$ SU(4) gates between connected pairs of qubits arranged in a grid. An analogous process is used to generate the linear connectivity circuits. (b) The average polarizations of the each type of random circuit, compiled for and run on \texttt{ibm\_hanoi}, estimated using our benchmark. The most significant differences in polarizations are seen in shape $(n,d)=(14,2)$ circuits and $(4,8)$ circuits. We use an exponential decay heuristic ($\bar{\gamma}_{n,d} = Ap^d$ for shapes with $d>1$ and $n \leq 10$, and  $\bar{\gamma}_{n,d} = Ap^n$ otherwise) to estimate the average polarization of circuits of shapes not included in our benchmark (light-outlined boxes).}
    \label{fig:grid_ibmq}
\end{figure}

Our benchmark generator does not require the high-level circuits that define the benchmark to have a particular structure (such as Haar-random two-qubit gates on random pairings of qubits, as is the case in QV circuits). Our benchmark generator can create random circuit benchmarks based on random circuits of any structure. This structure can be tailored to measure the performance of a processor in a desired setting---e.g., with a particular connectivity or gate set.
The all-to-all connectivity in QV circuits is not necessary for all applications, and implementing the large SWAP networks required to run QV circuits on processors without high connectivity greatly limits the shapes of QV circuits that can be used to usefully benchmark near-term quantum processors. Therefore, benchmarks based on other geometries are also potentially interesting and useful. In this section, we illustrate this application of our method by demonstrating scalable full-stack benchmarks, on \texttt{ibm\_hanoi}, based on random circuits defined on grid and linear geometries.

We construct the high-level random circuits for our benchmarks by sampling layers of Haar-random SU(4) gates restricted to a grid or linear connectivity graph. These two-qubit gates are sampled so that the average number of two-qubit gates per layer is $\nicefrac{n}{4}$. We ran these randomized grid and linear geometry benchmarks on \texttt{ibm\_hanoi}. We ran grid connectivity circuits with 17 different shapes and linearly connectivity circuits of 19 different shapes (Fig.~\ref{fig:grid_ibmq}, black boxes). The classical preprocessing method we we use in these demonstrations does approximation of the target unitary, and it adds X-X dynamical decoupling on idling qubits during two-qubit gates \footnote{We note that many compilations optimizations have been developed to improve processor performance and studied in the context of QV (\cite{cross2018validating,Baldwin2022reexaminingquantum}). We do use heavily optimized classical pre-processing here, and therefore our results are not expected to reflect the optimal full-stack processor performance.}.

Figure~\ref{fig:grid_ibmq} shows the average polarizations obtained from our randomized full-stack benchmarks (boxes with black outlines). Our results show little difference in performance between grid and linear geometry random circuits. The largest differences are in the $(n,d)=(14,2)$ circuits and the shape $(4,8)$ circuits. The shape $(4,8)$ grid geometry circuits, when compiled, have an average of $7.5$ more two-qubit gates than the linear geometry circuits, and therefore we expect lower polarizations. While the difference in average gate count increases with increasing depth (and width), most larger circuits of both circuit types have very low polarization ($\gamma \approx 0$), making the results indistinguishable. In contrast, the shape $(14,2)$ grid and linear geometry circuits are much closer in average two-qubit gate count (they differ by approximately $2$ two-qubit gates).

\section{Scalable Full-Stack Algorithm-Based Benchmarks}
\label{sec:alg_circuits}

\begin{figure}
    \centering
    \includegraphics{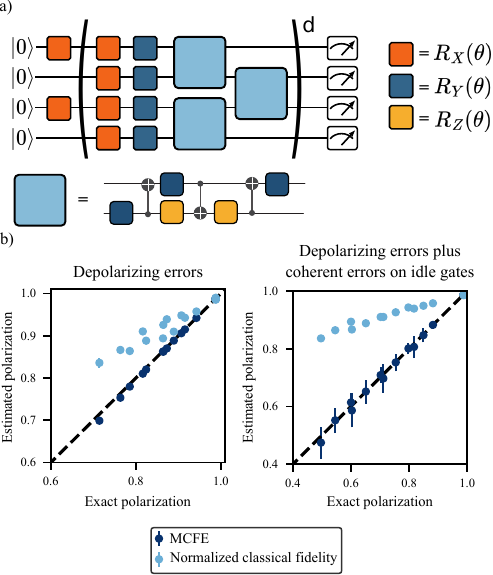}
    \caption{\textbf{A scalable Hamiltonian simulation full-stack benchmark.} We use our method to construct a benchmark from Hamiltonian simulation circuits. (a) The structure of the $n$-qubit Hamiltonian simulation circuits we used, for the case of $n=4$. Each circuit consists of a subroutine repeated $d$ times, and we call $d$ the benchmark depth of the circuit.  We use our benchmark generator to construct a scalable full-stack benchmark with Hamiltonian simulation circuits. (b) We simulated our benchmark with a local depolarizing error model (left plot) and a local depolarizing model with added coherent $Z$ error on idling qubits (right plot) for varied-shape circuits. We compare the polarizations estimated by our method to the exact circuit polarizations and the normalized classical fidelity between the ideal and observed base circuit output distributions. For both error models, the normalized classical fidelity overestimates the polarization, and the effect is larger in the error model with coherent $Z$ errors.}
    \label{fig:alg_sim}
\end{figure}

The benchmarks we have presented so far come from applying our benchmark generator to \emph{random circuits}. Our method can also be applied to structured circuits to create efficient and robust benchmarks that measure a full stack processor's ability to run quantum algorithms, or subroutines thereof. In this section, we use our benchmark generator to demonstrate full-stack benchmarking with circuits for Hamiltonian simulation. 

We applied our benchmark generator to Hamiltonian simulation circuits that are based on those from the Quantum Economic Development Consortium (QED-C)'s benchmarking suite, introduced in Ref.~\cite{lubinski2023applicationoriented}. These circuits simulate an anti-ferromagnetic chain of spins using a Trotterization of a target Hamiltonian. We take the benchmark depth $d$ of our circuits to be the Trotter order of the approximation, which corresponds to the number of repetitions of a particular subcircuit. Each Hamiltonian simulation circuit has parameters $h_z, h_x \in [-1,1]$ determining the Hamiltonian to be simulated. In Ref.~\cite{lubinski2023applicationoriented}, these are fixed to particular values (and the Trotter order is fixed). In our benchmark, we sample uniformly random values $h_z, h_x \in [-1,1]$ to create a set of high-level circuits. These circuits all have the same high-level gate structure but different gate parameters (see Fig.~\ref{fig:alg_sim}). 

We simulated our Hamiltonian simulation full-stack benchmark on a heavy-hexagon processor geometry  using a compilation routine that performs exact compilation (up to qubit permutation) of the target circuits. We ran these simulations with two error models: (1) a gate-independent depolarizing error model, with single-qubit gate polarization of $\gamma_{1Q} = 0.999$ and two-qubit gate polarization of $\gamma_{2Q}=0.99$, and (2) the same depolarizing error as in (1), with the addition of $Z$-axis rotation error on each idling qubit during two-qubit gates, where each idling qubit experiences a $Z$-axis rotation of $0.08$ radians. 

We also compare the results of our benchmark to the results obtained by running the compiled circuits and then applying the analysis specified in Ref. \cite{lubinski2023applicationoriented}. That analysis consists of computing the normalized classical fidelity between the ideal and observed output distributions \cite{lubinski2023applicationoriented},
\begin{equation}
    \tilde{F} = \frac{F(p_{\textrm{ideal}}, p_{\textrm{observed}})-\sum_{x \in \{0,1\}^n} \frac{1}{2^n}\sqrt{p_{\textrm{ideal}}(x)}}{1-\sum_{x \in \{0,1\}^n} \frac{1}{2^n}\sqrt{p_{\textrm{ideal}}(x)}}, \label{eqn:cf}
\end{equation}
where $\{p_{\textrm{ideal}}(x)\}_{x \in \{0,1\}^n}$ is the ideal output distribution of the circuits and $\{p_{\textrm{observed}}(x)\}_{x \in \{0,1\}^n}$ is the observed output distribution. Computing the classical fidelity requires the output distribution for the error-free circuit, necessitating full simulation of each high-level circuit. Furthermore, Eq.~\eqref{eqn:cf} is ill-conditioned when $p_{\textrm{ideal}}$ is close to the uniform distribution. As a result, the circuits in \cite{lubinski2023applicationoriented} are chosen so that the error-free output distribution of the circuits is strongly peaked at a small set of outcomes, which is not necessarily reflective of the output distribution for typical applications.

Figure~\ref{fig:alg_sim} shows the results of our simulations. Figure~\ref{fig:alg_sim}a compares the results of our benchmark with the exact circuit polarizations.  We observe close agreement between the results of our benchmarks and the exact polarizations of the compiled circuits. In contrast, the normalized classical fidelity is systematically higher than the exact polarization and the results of our benchmark. We conjecture that this is because the classical fidelity is only sensitive to changes in the classical output distribution of the compiled circuit, but some errors do not cause changes to the output distribution (whereas they \emph{do} decrease the process fidelity). In Appendix~\ref{app:fidelities}, we show that for a simplified error model and definite-outcome circuits, the normalized classical fidelity is greater than the polarization under broad conditions. The discrepancy between the normalized classical fidelity and the polarization is even larger in the presence of coherent $Z$ errors on idle gates, which is because the classical fidelity is not sensitive to $Z$ errors preceding a measurement.  Furthermore, there is significant variance in the polarizaion estimates from our benchmark. These results demonstrate that gate parameters, not just circuit structure, can significantly impact performance, as  the way that errors propagate depends on the gate parameters. It is therefore useful for benchmarks like the ones presented here to include varied gate parameters.

\begin{figure}
    \centering
    \includegraphics{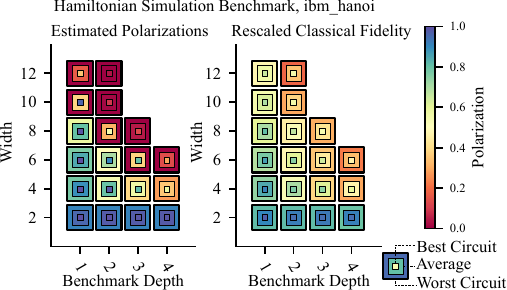}
    \caption{\textbf{Scalable full-stack Hamiltonian simulation benchmarking on IBM Q.} We ran our scalable Hamiltonian simulation full-stack benchmark on \texttt{ibm\_hanoi}. We compare the results of our benchmark (left plot) to the results of running the compiled circuits and measuring the classical fidelity of the output distribution (right plot), which we rescale to an effective polarization using the approach from Ref.~\cite{lubinski2023applicationoriented}. We find that our benchmark indicates much lower polarization than the classical fidelity analysis with $n>2$ qubits. We also observe large variations in the polarization among circuits of the same shape, suggesting the presence of coherent error.}
    \label{fig:alg_ibmq}
\end{figure}

We ran our Hamiltonian simulation full-stack benchmark on \texttt{ibm\_hanoi}, using a compilation routine that performs exact compilation (up to qubit permutation) and adds X-X dynamical decouping on idling qubits during two qubit gates. Figure~\ref{fig:alg_ibmq} shows the results of our benchmark, compared to the normalized classical fidelity [Eq.~\eqref{eqn:cf}] calculated from the observed output distribution of each compiled circuit. The polarizations estimated by our method are consistently lower than the normalized classical fidelity for $n > 2$, which is consistent with the results of our simulations. We conjecture that the normalized classical fidelity is lower than the polarizations estimated by our benchmark for $n=2$ due to state preparation and readout error---the classical fidelity of the compiled circuits is sensitive to the quantum processor's readout error, whereas our benchmark is, by design, not sensitive to readout error (although it can be used to separately estimate readout error).
We observe a rapid decrease in polarization with increasing width. The normalized classical fidelity decreases more gradually with width---the mean normalized classical fidelity of shape $(12, 1)$ circuits is approximately $0.57$, whereas the mean polarization estimated by our benchmark is approximately $0.17$. 
The polarization and rescaled classical fidelity also decrease rapidly with increasing benchmark depth for $n>2$. The number of two-qubit gates in the compiled circuits increases significantly with depth for $n>2$---for example, our shape $(6,2)$ compiled circuits have $30$ two-qubit gates, and each additional benchmarking layer adds $15$ two-qubit gates to the width $n=6$ circuits. In contrast, each two-qubit circuit gets compiled (without approximation) into a circuit with exactly $3$ two-qubit gates, allowing the processor to perform all of these circuits with high fidelity.

\section{Discussion}

Precise and principled benchmarking of the full quantum stack enables reliable assessment of progress towards useful quantum computation. In this paper, we have introduced a general method for creating computationally efficient full-stack benchmarks from any set of invertible circuits. Our method avoids the exponentially expensive classical simulations found in existing full-stack benchmarks by using compiled circuits output by the classical preprocessor to build  efficiently verifiable benchmarking circuits. We have demonstrated how our method can turn existing benchmarks---namely, the quantum volume benchmark and the QED-C's application-oriented benchmarks---into efficient and robust full-stack benchmarks. However, our method is applicable to any set of unitary circuits, and we anticipate it being used to construct scalable benchmarks from other interesting families of circuits.

We have demonstrated three applications of our benchmark generator, demonstrating its versatility, and we have shown that its results are a precise measure of processor performance. As demonstrated by our MQV protocol, our method enables efficient average performance benchmarking of large quantum processors by eliminating the need for expensive classical circuit simulation. Additionally, our benchmarks measure process fidelity and are therefore sensitive to errors that do not affect computational basis measurement outcomes, allowing them to be used to accurately assess a processor's performance on subroutines that will not be directly followed by measurements. Using variable-shape circuits, we demonstrated the use of our full-stack randomized benchmarks to generate detailed volumetric performance data. In the near term, our technique can be used to assess progress towards useful quantum computation, both using randomized benchmarks with varied circuit sampling and through application-based benchmarks based on important subroutines for quantum algorithms. In the long term, we anticipate our method being used to create precision benchmarks to efficiently assess large processors' performance on full quantum algorithms.

\section*{Acknowledgements}
This material was funded in part by the U.S. Department of Energy, Office of Science, Office of Advanced Scientific Computing Research, Quantum Testbed Pathfinder Program, and by the Laboratory Directed Research and Development program at Sandia National Laboratories. T.P.~acknowledges support from an Office of Advanced Scientific Computing Research Early Career Award. Sandia National Laboratories is a multi-program laboratory managed and operated by National Technology and Engineering Solutions of Sandia, LLC., a wholly owned subsidiary of Honeywell International, Inc., for the U.S. Department of Energy's National Nuclear Security Administration under contract DE-NA-0003525. All statements of fact, opinion or conclusions contained herein are those of the authors and should not be construed as representing the official views or policies of the U.S. Department of Energy, or the U.S. Government. We acknowledge the use of IBM Quantum services for this work. The views expressed are those of the authors, and do not reflect the official policy or position of IBM or the IBM Quantum team.

\bibliography{main}

\appendix

\subsection{Method}
\label{app:method}
In this appendix, we provide further details on our full-stack benchmarking method. The circuits our benchmark generator creates are forms of mirror circuits \cite{proctor2020measuring}. Our benchmarks use mirror circuit fidelity estimation (MCFE) \cite{proctor2022establishing} to estimate the fidelity of each compiled circuit. Our method's efficiency arises from two features of MCFE:
\begin{enumerate}
    \item \emph{Gate-efficient fidelity estimation:} MCFE uses a gate-efficient form of fidelity estimation with local unitary 2-designs to reliably measure the fidelity of a channel that ideally implements the identity. This requires adding only two layers of single-qubit gates to the circuits, for local state preparation and measurement.
    \item \emph{Circuit mirroring:} Because the compiled quantum circuits we'd like to run in practice are not identity circuits, our method runs mirror circuits containing the compiled circuits of as subroutines. These mirror circuits ideally implement a Pauli, making their outputs easily verifiable, and they only require local computations (i.e., inverting individual single- and two-qubit gates) to construct. Therefore, no expensive classical computation is required to construct or verify the output of our benchmarking circuits, making our method scalable.
\end{enumerate} 

The mirror circuits our method uses involve the circuit output by the classical preprocessor along with added subroutines. Mirror circuit fidelity estimation ensures robustness to error in added subcircuits using randomized compilation \cite{wallman2015noise} on added subcircuits to twirl their error into stochastic error.

We now describe the circuits our method uses to estimate the fidelity of the processor's implementation of $\comp{U}{C}$ [$\Lambda$] to the target unitary evolution $\mathcal{U}'$, where $\mathcal{U}' = \mathcal{P}\mathcal{U}$. We require an exact compilation $\exact{U}{C}$ of $\mathcal{U'}$ that contains only single-qubit gates and self-inverse Clifford two-qubit gates. We will also use circuit layers $\gate{L}$ consisting of gates sampled from  $\otimes_{i=1}^n \mathcal{L}_i$, where $\mathcal{L}_i$ is a single-qubit unitary 2-design. For a circuit $\gate{C}$, let $\gate{C}_{\textrm{rev}}$ denote its layer-by-layer inverse. We use $f_{\textrm{rc}}(\gate{G})$ to denote a randomized compilation of a circuit $\gate{G}$.  We generate $k \gg 1$ of each of the following three types of mirror circuits:
\begin{enumerate}
\item Circuits containing the compiled circuit output by the classical preprocessor and the inverse of an exact compilation:
\begin{equation}
    M_1 =  f_{\textrm{rc}}(\gate{L}) \comp{U}{C}  f_{\textrm{rc}}(\exact{U}{C}_{\textrm{rev}}\gate{L}_{\textrm{rev}})
\end{equation}
\item Circuits with the exact compilation and its inverse:
\begin{equation}
    M_2 = f_{rc}(\gate{L}\comp{U}{C}\exact{U}{C}_{\textrm{rev}}\gate{L}_{\textrm{rev}})
\end{equation}
\item State preparation and measurement circuits: 
\begin{equation}
    M_3 = f_{\textrm{rc}}(\gate{L}\gate{L}_{\textrm{rev}})
\end{equation}
\end{enumerate}
Each mirror circuit, when implemented perfectly, will always output a single bit string, which we call its \emph{target bit string}. We use the results of running of these three types of circuits to estimate the fidelity of $F(\Lambda,\mathcal{U})$. For each mirror circuit $m$, we compute its \emph{observed polarization}, 
\begin{equation}
\gamma(m) =  \frac{4^n}{4^n-1}\left[\sum_{k=0}^{n} \left(-\frac{1}{2}\right)^k h_k\right] - \frac{1}{4^n -1},
\label{eq:appS}
\end{equation}
where $h_k$ is the probability that the circuit outputs a bit string with Hamming distance $k$ from its target bit string. 
We have, upon averaging over randomized compilations \cite{proctor2022establishing},
\begin{align}
	\mathbb{E}[\gamma(M_1)] & \approx \gamma(\gate{L})\gamma(\comp{U}{C})\gamma(\exact{U}{C}_{\textrm{rev}})\gamma(\gate{L}_{\textrm{rev}}) \label{eq:m1} \\
	\mathbb{E}[\gamma(M_2)] & \approx \gamma(\gate{L})\gamma(\exact{U}{C})\gamma(\exact{U}{C}_{\textrm{rev}})\gamma(\gate{L}_{\textrm{rev}}) \\
	& \approx\gamma(\gate{L})\gamma(\gate{L}_{\textrm{rev}})\gamma(\exact{U}{C}_{\textrm{rev}})^2 \label{eq:m2}\\
	\mathbb{E}[\gamma(\bar{M}_3)] & \approx \gamma(\gate{L})\gamma(\gate{L}_{\textrm{rev}}), \label{eq:m3}
\end{align}
which results in the equation for polarization shown in Figure~\ref{fig:protocol}. We use the results of our sampled circuits to estimate the expected polariations given in Eqns.~\eqref{eq:m1}-~\eqref{eq:m3}. Our estimate of the fidelity of $\Lambda$ is
\begin{equation}
\hat{F}(\Lambda) \approx 1 - \frac{4^n-1}{4^n}\left(1 - \frac{\avg[\gamma(M_1)]}{[\avg(M_2)][\avg(M_3)]}\right).
\end{equation}

We make one further modification to MCFE to slightly reduce the gate count of our benchmarking circuits --- rather than creating an exact compilation of $\mathcal{U}'$ and inverting it to obtain $\gate{C}_{\textrm{rev}}$, we create an exact compilation of $\mathcal{P}'\mathcal{U'}^{-1}$ for some permutation $\mathcal{P}'$, and use this circuit as $\gate{C}_{\textrm{rev}}$ (and its layer-by-layer inverse as $\gate{C}$). To compensate for this effect, in the $M_1$ circuits the single-qubit gates in $\gate{L}_{\textrm{rev}}$ are permuted according to $\mathcal{P}'$, so that $\gate{L}\comp{U}{C}\exact{U}{C}_{\textrm{rev}}\gate{L}_{\textrm{rev}}$ implements a SWAP network if $\comp{U}{C}$ is an exact compilation of $\mathcal{U}$. This modification slightly reduces the number of two-qubit gates in our mirror circuits.

Fidelity estimation with local 2-designs is not the only form of efficient fidelity estimation: direct fidelity estimation requires only single-qubit state preparation and measurement, and is computationally efficient on circuits that implement a Clifford gate \cite{flammia2011direct, dasilva2011practical}. Our technique can therefore be adapted to use mirror circuits that only invert $\mathcal{U}'$ up to any Clifford. However, we do not explore this option here.

\subsection{Mirror quantum volume}
\label{app:mqv}

\begin{figure}
    \centering
    \includegraphics{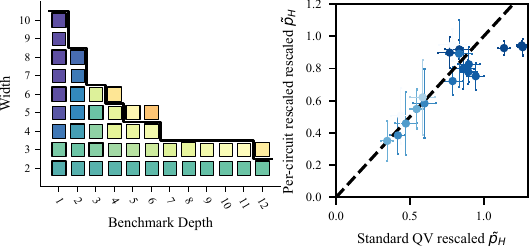}
    \caption{\textbf{Rescaling the heavy output probability of QV circuits.} We compare the per-circuit rescaling of the QV heavy output probabilities in Section~\ref{sec:mqv_demo} to an aggregate rescaling based on the standard QV benchmark. These analyses agree on high-depth circuits (lighter dots in the plot on the right). The standard QV rescaling results in significantly higher values for high-width, low-depth circuits (darker points in the plot on the right ).}
    \label{fig:app_qv}
\end{figure}
Eq.~\eqref{eq:tilde_ph} is not guaranteed to reliably estimate the heavy output probability for very shallow circuits due to large deviations in the outcome statistics from the Porter-Thomas distribution---i.e. for ideal, shallow QV circuits, it is not true that $p_H \approx \nicefrac{1}{3\ln 2}$ .  Eq.~\eqref{eq:tilde_ph} can be modified to accurately reflect the heavy output probability of circuits of a particular shape by rewriting Eq.~\eqref{eq:tilde_ph} as 
\begin{equation}
    \gamma = \frac{\tilde{p}_H - \frac{1}{2}}{p_H - \frac{1}{2}}. \label{eq:tilde_gamma}
\end{equation}
We can estimate the ideal average heavy output probability of depth $d$, width $n$ QV circuits via simulation, and use this value as $p_H$. Note, however, that this calculation is only feasible in the regime where classical simulation of the QV circuits is feasible. In our anaysis of the \texttt{ibm\_hanoi} data in Section~\ref{sec:mqv_demo}, we use Eq.~\eqref{eq:tilde_gamma} on a per-circuit basis---i.e., we let $p_H$ be the ideal heavy output probability for a single circuit and $\tilde{p}_H$ be the observed heavy output probability for that circuit, and we estimate $\gamma$ separately for each QV circuit. 

Figure~\ref{fig:app_qv} shows the results of applying the standard QV rescaling to our QV results from \texttt{ibm\_hanoi}---i.e., setting $p_H = (1+\ln{2})/2$ regardless of circuit shape. This rescaling results in significantly greater rescaled heavy output probabilities than the per-circuit analysis for high $n$, low $d$ circuits. These rescaled heavy output probabilities are greater than 1 for the most shallow circuits, which happens because these circuits typically have ideal heavy output probability much greater than $(1+\ln{2})/2$ For high-depth circuits, the results of the two analyses are approximately the same.

\subsection{Classical fidelity and process fidelity}
\label{app:fidelities}

In this appendix, we show that, the normalized classical fidelity [Eq.~\eqref{eqn:cf}] upper bounds the process fidelity for a simple error model and type of circuit. We assume that the ideal circuit is a definite outcome circuit, i.e., it outputs a single bit string. We will model the circuit's error by a single \emph{overall error channel} $\mathcal{E}$---i.e., we let $\Lambda = \mathcal{E}\mathcal{U}$---and we will consider two different models for $\mathcal{E}$. First, we will assume that $\mathcal{E}$ is a tensor product of single-qubit depolarizing channels, each with polarization $\gamma$. The normalized classical fidelity of the process $\Lambda$, with respect to the ideal process $\mathcal{U}$, is
\begin{equation}
\tilde{F} = \frac{1}{2^n-1}\left((1+\gamma)^n - 1\right), \label{eq:depol_cf}
\end{equation}
and the process fidelity is \begin{align}
    F & = \frac{1}{4^n-1}(1+3\gamma)^n-\frac{1}{4^n-1} \\
    & = \frac{1}{4^n-1}\sum_{k=0}^n{n \choose k}(1+\gamma)^{k}\gamma^{n-k}-\frac{1}{4^n-1} \\
    & \leq \frac{1}{4^n-1}\left((1+\gamma)^n + \sum_{k=1}^n{n \choose k}(1+\gamma)^k\right)-\frac{1}{4^n-1} \\
    & \leq \frac{1}{4^n-1}\left((1+\gamma)^n-1\right) + \sum_{k=1}^n{n \choose k}(1+\gamma)^k \label{eq:depol_pf}
\end{align} 
We will now show that $\tilde{F}  -F \geq 0$. Combining Eq.~\eqref{eq:depol_cf} and Eq.~\eqref{eq:depol_pf}, we have
\begin{align}
    \tilde{F}-F & \geq \frac{2^n}{4^n-1}\left((1+\gamma)^n - 1\right) + \frac{1}{4^n-1}\sum_{k=1}^n{n \choose k}(1+\gamma)^k \\
    & \geq 0.
\end{align}
Therefore, the classical fidelity upper bounds the process fidelity when $\mathcal{E}$ is a tensor product of single-qubit depolarizing error channels. 

We now consider a slightly more general model, where the overall error channel is an $n$-qubit stochastic Pauli channel, i.e. $\mathcal{E} = \sum_{\mathcal{P} \in \mathbb{P}_n} \gamma_{\mathcal{P}} \mathcal{P}$, where $\mathbb{P}_n$ denotes the set of $n$-qubit Pauli superoperators. The process fidelity of a stochastic Pauli channel is $\gamma_{\Id_n}$, where $\Id_n$ denotes the $n$-qubit identity operator. The polarization of $\mathcal{E}$ is 
\begin{equation}
    \gamma = \frac{4^n}{4^n-1}\gamma_{\Id_n} - \frac{1}{4^n-1}.
\end{equation}
We call a Pauli $\mathcal{P} \neq \mathbb{I}_n$ that is a tensor product of only $Z$ and $I$ Pauli superoperators a \emph{Z-type Pauli} superoperator, and we let $\mathbb{Z}_n$ denote the set of $n$-qubit $Z$-type Pauli superoperators. The classical fidelity is insensitive to Z-type Pauli errors in the overall error channel. The normalized classical fidelity is 
\begin{equation}
    \tilde{F} = \frac{2^n}{2^n-1}\left(\gamma_{\Id_n} + \sum\limits_{\mathcal{P} \in \mathbb{Z}_n}\gamma_{\mathcal{P}}-\frac{1}{2^n}\right).
\end{equation}
The difference between the normalized classical fidelity and the polarization is therefore
\begin{align}
    \tilde{F} - \gamma & =  \left(\frac{1}{2^n-1} - \frac{1}{4^n-1}\right)\left(\gamma_{\Id_n}-1\right) + \frac{2^n}{2^n-1}\sum\limits_{\mathcal{P} \in \mathbb{Z}_n}\gamma_P.
\end{align}
Therefore, $\tilde{F}-\gamma \geq 0$ when
\begin{align}
    \sum\limits_{\mathcal{P} \in \mathbb{Z}_n}\gamma_P & \geq \left(\frac{1}{2^n} - \frac{2^n-1}{4^n-1}\frac{1}{2^n}\right)\left(1-\gamma_{\Id_n}\right) \\
    & \geq \frac{2^n-1}{4^n-1}\left(1-\gamma_{\Id_n}\right). 
\end{align}
Informally, the condition above is satisfied when the rate of $Z$-type errors is at least at the rate of $Z$-type errors in a global depolarizing channel with the same process fidelity as $\mathcal{E}$. This means that in systems where errors are biased towards Z-type errors (which is common in current quantum processors based on superconducting qubits), we expect the classical fidelity to overestimate the process fidelity. 
\end{document}